\documentclass[superscriptaddress,showpacs,floatfix,prl,twocolumn]{revtex4}
\usepackage{amssymb}

\usepackage{graphicx}

\newcommand{\be}{\begin{equation}}
\newcommand{\ee}{\end{equation}}

\newcommand{\kk}{{\mathbf k}}

\begin{document}

\title{Excitations in a non-equilibrium Bose-Einstein condensate of
  exciton-polaritons} 
\author{Michiel Wouters}
\affiliation{TFVS, Universiteit Antwerpen, Groenenborgerlaan 171, 2020
  Antwerpen, Belgium} 
\author{Iacopo Carusotto}
\affiliation{BEC-CNR-INFM and Dipartimento di Fisica, Universit\`a di
  Trento, I-38050 Povo, Italy}

\begin{abstract}
We have developed a mean-field model to describe the dynamics of a
non-equilibrium Bose-Einstein condensate of exciton-polaritons in a
semiconductor microcavity. The spectrum of elementary excitations
around the stationary state is  analytically studied in different
geometries. A diffusive behaviour of the Goldstone mode is found in
the spatially homogeneous case and new features are predicted for the
Josephson effect in a two-well geometry.
\end{abstract}
\maketitle

After a few decades of impressive efforts on a variety of different
systems such as bulk cuprous oxide~\cite{Cu2O} and coupled quantum
wells~\cite{CQW}, first observations of Bose-Einstein condensation
(BEC) of excitons in solid state systems have been recently reported
in a gas of exciton-polaritons~\cite{kasprzak} and immediately
confirmed by other groups~\cite{Yamamoto,GaN}. The system under
investigation consists of a semiconductor microcavity containing a few
quantum wells with an excitonic transition strongly coupled to the
cavity photon mode. In this strong coupling regime, the basic
excitations of the system are exciton-polaritons, i.e. linear
superpositions of a quantum well exciton and a cavity photon. As
compared to other examples of BEC, namely in liquid $^4$He and
ultracold atomic gases, the main novelty of the present polariton
system is its intrinsic non-equilibrium nature due to the finite
lifetime of polaritons. The condensate has in fact to be continuously
replenished from the relaxation of optically injected high energy
excitations (e.g. free carriers or hot polaritons), and its steady
state results from a dynamical equilibrium between pumping and losses.
This makes the present system a unique candidate for the study of the
BEC phase transition in a non-equilibrium context.  Recent
theoretical work~\cite{littlewood} has suggested that the
non-equilibrium condition is responsible for dramatic changes in
the dispersion of low-lying excitations of incoherently pumped
polariton condensates: the sound mode of equilibrium condensates is
replaced by a diffusive mode with flat dispersion, as it typically
happens in coherently driven pattern forming systems, such as Benard
cells in heat convection~\cite{cross} or optical parametric
oscillators~\cite{goldstone}.  

The present Letter is devoted to the development of a simple and generic
model of a non-equilibrium condensate which does not involve the
microscopic physics of the polariton, and can be used to describe the
dynamics independently of the details of the specific pumping scheme. Our
model is inspired by classical treatments of laser operation~\cite{laser},
and closely resembles the generic model of atom lasers developed in~\cite
{atomlaser}. In this way, we are able not only to confirm the conclusions of
Ref.~\cite{littlewood} but also to analytically relate the elementary
excitation spectrum to experimentally accessible quantities. The same model
is then applied to the Josephson effect~\cite{Josephson,smerzi,oberthaler} 
in a system of two weakly coupled polaritonic condensates:
 predictions are given 
for the frequency and the intrinsic damping rate of Josephson oscillations, and 
overdamped behavior  is anticipated in the case of strong damping.

\begin{figure}[htbp]
\begin{center}
\includegraphics[width=1.\columnwidth,angle=0,clip]{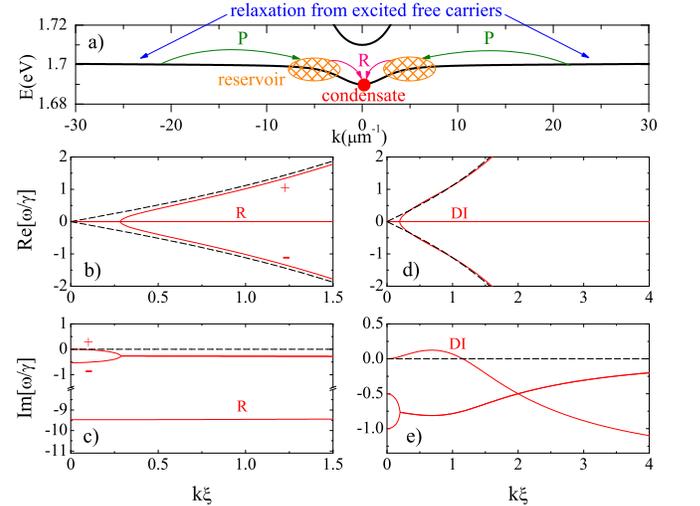}
\end{center}
\caption{Top (a) panel: sketch of the pumping and condensate
  formation scheme. (b-e) panels: real and imaginary part of the
  excitation spectrum of a homogeneous polariton condensate as a
  function of the wavevector $k$ (in units of the healing length
  $\protect\xi=\protect\sqrt{\hbar/(m_{LP}\,\protect\mu)}$).
  Parameters: $\protect\gamma_{R}/\protect\gamma=5$ and
  $\protect\alpha=\protect\beta=\protect\gamma/\protect\mu=1$ (b,c);
  $\protect\gamma_{R}/\protect\gamma=1$ and $\protect\alpha=0.5, \protect\beta=
  \protect\gamma/\protect\mu=\gamma_R/\mu=1$ (d,e). Rescaled diffusion
  constant $D\,m_{LP}/\hbar\approx 5\times 10^{-4}$; the spectra are
  indistinguishable from the $D=0$ case. Dashed lines in (b,d):
  Bogoliubov dispersion (\ref{bog_eq}) at equilibrium.}
\label{fig:disp}
\end{figure}

The experimental scheme used to create the polariton condensate is
sketched in Fig.\ref{fig:disp}a: under a continuous-wave high energy
illumination, hot free carriers are generated in the semiconductor
material forming the microcavity.  Their cooling down by phonon
emission leads the formation of a incoherent gas of bound excitons in
the quantum wells, which eventually accumulate in the so-called
bottleneck region above the inflection point of the lower
exciton-polariton (LP) branch~\cite{tassone_bottleneck}.
Polariton-polariton collisions are then responsible for the (generally
slower) scattering of polaritons from the bottleneck region to the
bottom of the LP branch. For high enough polariton density, Bose
stimulation of scattering into the lower part of the LP can take
place~\cite{porras}. When the stimulated scattering rate overcomes
losses, the polariton field becomes coherent, and a Bose-Einstein
condensate appears~\cite{note1}.

Our model is based on a mean-field description of the coherent
polariton field in terms of a generalized Gross-Pitaevskii equation
(GPE) for the condensate macroscopic wavefunction including loss 
and amplification terms
\begin{equation}
i\frac{\partial \psi}{\partial t}=\left\{ -\frac{\hbar \nabla ^{2}}{2m_{LP}}+
\frac{i}{2}\big[R(n_R)-\gamma\big]+g\,|\psi|^{2}+2\tilde{g}\,n_{R}\right\}
\psi .  \label{eq:GP}
\end{equation}
As we are interested in the lowest part of the LP dispersion, a
parabolic dispersion can be taken for the polaritons, with an
effective mass $m_{LP}$.  The strength of polariton-polariton
interactions within the condensate is fixed by the coupling constant
$g$, while $\gamma$ is the polariton damping rate at the bottom of the
band.  Assuming that relaxation processes are fast enough to ensure
local equilibrium in the polariton reservoir in the bottleneck region,
and that all coherences between the reservoir and the condensate decay
on a fast time-scale compared to the condensate dynamics, the state of
the reservoir is fully determined by its polariton density $n_R(x)$.
The amplification rate of the condensate due to stimulated scattering
of polaritons from the reservoir is a monotonically growing
function $R\left(n_{R}\right)$ of the reservoir density
$n_R$. Interactions between the condensate and the reservoir
polaritons are modelled by the interaction constant $\tilde{g}$,
generally different from the condensate one $g$.  With respect to
single-mode theories~\cite{4pani,otherThPolBEC}, the generalized GPE
has the important advantage of fully taking into account the
multi-mode nature of the spatially extended polariton
field. Moreover, unlike kinetic approaches based on the Boltzmann
equation~\cite{tassone,porras,haug}, the present model is able to
describe the coherent dynamics of the condensate. These issues are
essential in view of a study of the elementary excitations of the 
condensate.

The evolution equation (\ref{eq:GP}) for the condensate has to be coupled to an
equation for the density $n_R(x)$ of reservoir polaritons: 
\begin{equation}
\frac{\partial n_R}{\partial t}=P-\gamma _{R}\,n_{R}-R\left(
n_{R}\right) \left\vert \psi \left( x\right) \right\vert ^{2}+
D\nabla ^{2}n_{R}.  \label{eq:rate_B}
\end{equation}
Polaritons are pumped in the reservoir with a rate $P$ and relax at a
rate $\gamma_R$. The spatial hole-burning effect due to the scattering
of reservoir polaritons into the condensate is taken into account by the 
$R(n_{R})\,|\psi|^2$ term; spatial diffusion of reservoir polaritons takes place
with a diffusion constant $D$.

The steady state of the system under a continuous-wave and uniform
pumping $P$ can be obtained by substituting the ansatz 
\begin{eqnarray}
\psi \left( x,t\right) &=&e^{-i\mu_T t}\,\psi _{0}  \label{eq:stat1} \\
n_{R}\left(x,t\right)&=&n_{R}^{0}.  \label{eq:stat2}
\end{eqnarray}
into (\ref{eq:GP}) and (\ref{eq:rate_B}). For small values of $P$,
no condensate is present $\psi_0=0$ and the reservoir density is a
linear function of the pump intensity $n_R^0=P/\gamma_R$. This
solution is dynamically stable as long as the amplification rate
is not able 
to overcome the losses, i.e. $R(n_R^0)<\gamma$. The threshold
$P=P_{th}$ corresponds to the value $n_R^{th}$ for the reservoir
density, which guarantees equilibrium between amplification and losses
$R(n_R^{th})=\gamma$. When the pumping rate $P$ is increased above the
threshold, the solution $\psi_0=0$ becomes dynamically unstable and a
condensate appears. Stationarity imposes the net gain to vanish, which
fixes the reservoir density to the equilibrium value
$n_R^0=n_R^{th}$. The condensate density grows as 
$n_{c}^{0}=\psi_{0}|^{2}=(P-P_{th})/\gamma$, and the oscillation
frequency of the macroscopic wavefunction is $\mu_T=\mu+2{\tilde
  g}n_R^0$ with $\mu=g\,n_c^0$.

As usual~\cite{bec-book,PRL_superfl,goldstone}, the elementary
excitations spectrum around the stationary state of the system can be
obtained by a linearization of the motion equations
(\ref{eq:GP}-\ref{eq:rate_B}) around the steady state solution
(\ref{eq:stat1}-\ref{eq:stat2}). Thanks to the translational
invariance of the system, the fluctuations can be decomposed in their
Fourier components: 
\begin{eqnarray}
\psi \left( \mathbf{r},t\right) &=&  e^{-i\mu_T t}\,\psi_{0}\, 
\Big[1+\sum_{{\mathbf{k}}}u_{{\mathbf{k}}}\,e^{i\left( {\mathbf{k}}\mathbf{r}-\omega
t\right)} + \nonumber \\
&+&v_{{\mathbf{k}}}^{\ast }\,e^{-i\left({\mathbf{k}}\mathbf{r}-\omega
t\right) }\Big], \\
n_{R}\left( t\right) &=&n_{R}^{0}\,
\left( 1+w_{{\mathbf{k}}}\,e^{i\left( {\mathbf{k}}\mathbf{r}-\omega t\right) }+
w_{{\mathbf{k}}}^{\ast}\,e^{-i\left( {\mathbf{k}}\mathbf{r}-\omega t\right) }\right).
\end{eqnarray}
Introducing the fluctuation vector $\mathcal{U}_{{\mathbf{k}}}=
\left(u_{{\mathbf{k}}},v_{{\mathbf{k}}},w_{{\mathbf{k}}}\right) ^{T}$ and substituting
this expansion into the motion equations, one is immediately led to the
eigenvalue equation $\mathcal{L}_{{\mathbf{k}}}\,
\mathcal{U}_{{\mathbf{k}}}=\omega\,\mathcal{U}_{{\mathbf{k}}}$ defining the elementary excitations,
where the generalized Bogoliubov matrix ${\mathcal{L}}_{\mathbf{k}}$ is
\begin{equation}
\mathcal{L}_{\kk}\!=\!\left( 
\begin{array}{ccc}
\mu+\frac{\hbar k^{2}}{2m_{LP}} & \mu & \frac{i\beta\,\gamma
}{2}+\frac{2\gamma\,\mu }{\alpha \gamma _{R}} \\
-\mu & -\mu-\frac{\hbar k^{2}}{2m_{LP}} & \frac{i\beta
  \,\gamma}{2}-\frac{2\gamma\,\mu }{\alpha \gamma _{R}} \\  
-i\alpha\,\gamma _{R}& -i\alpha\,\gamma _{R} & -i\left[ \eta\,\gamma
    _{R}+D\,k^{2}
\right]
\end{array}
\right)  \label{eq:bog}
\end{equation}
Here $\alpha = P/P_{th}-1$ is the relative deviation from the
threshold pumping intensity and the dimensionless coefficient $\beta
=n_{R}^{0}R^{\prime }\left( n_{R}^{0}\right) /R\left(n_{R}^{0}\right)$
characterizes the dependence of the amplification rate on the
reservoir density, and $\eta=1+\alpha\beta$.  The standard
Hartree-Fock value ${\tilde g}=2g$ has been taken. Quite remarkably,
the excitation spectrum does not depend on the actual value of the
scattering rate $R$ of the reservoir into the condensate mode: only
the effective exponent $\beta$ and the relative pumping rate $\alpha$
do matter.  Of course, the threshold value $P_{th}$ of the pumping
intensity does depend (in an inversely proportional way) on $R$.

Typical examples of the dispersion of elementary excitations around the
Bose condensed state are shown in Fig. \ref{fig:disp}(b-d). As the onset of Bose
condensation corresponds to a spontaneous breaking of a $U(1)$ symmetry, all
spectra involve a Goldstone branch whose dispersion $\omega_G(k)
$ tends to $0$ in the long-wavelength $k\rightarrow 0$ limit~\cite{bec-book,cross}. 
Physically, this mode can be understood as a slow rotation of the condensate phase
across the sample; the generator $\left( 1,-1,0\right)^{T}$ of global phase
rotations is indeed an eigenvector of $\mathcal{L}_{{\mathbf{k}}=0}$ with a
vanishing eigenvalue.

Let us analyze the different cases in more detail, starting from the
physically most relevant one $\gamma _{R}\gg \gamma$ where the state
of the reservoir always remains very close to its stationary-state for
the given density of condensate polaritons.  The dispersion for this
case is shown in Fig.\ref{fig:disp}(bc): in stark contrast with the
linear dispersion of the propagating sound mode in equilibrium
Bose-Einstein condensates~\cite{bec-book,note2}, the Goldstone mode
(indicated as $+$ in the figure) shows here a diffusive and
non-propagating behavior at low $k$. 
The real part is dispersionless and equal to zero, while the imaginary
part starts from zero in a quadratic way.  This result is in agreement
with the prediction of the recent work~\cite{littlewood} where these
issues were addressed starting from a very specific microscopic model
of the polaritons, and suggests that the diffusive behaviour of the
Goldstone mode is indeed a generic fact in non-equilibrium phase
transitions not only under a coherent pumping as in pattern forming
systems~\cite{cross,goldstone}, but also in the case of incoherent
pumping. Note that this diffusive behaviour is in no way due to the
spatial diffusion of reservoir polaritons, and would be present even
in the absence of spatial diffusion. The value
$D=5\,\mathrm{cm}^{2}/\mathrm{s}$ actually chosen in the figures is
inspired by recent experimental studies on CdTe quantum well
samples~\cite{portella}, and corresponds to a very small value of the
dimensionless diffusion constant ${\tilde D}=D\,m_{LP}/\hbar\approx
5\times 10^{-4}$.

An analytical explanation of this behavior is readily obtained by
adiabatically eliminating the dispersionless and strongly damped
reservoir mode ($R$ in the figure) whose imaginary part is close to
$(1+\alpha\beta)\,\gamma_R$.  Taking for simplicity $D=0$, this leads
to the following dispersion of the two branches of condensate
excitations:
\begin{equation}
\omega_\pm(k) =-\frac{i\Gamma}{2}\pm \sqrt{\omega _{Bog}(k)^2 -
\frac{\Gamma^{2}} {4}},  \label{spectr}
\end{equation}
where $\omega_{Bog}$ is
the usual Bogoliubov dispersion of dilute Bose gases at equilibrium 
\begin{equation}
\omega_{Bog}(k)=\sqrt{\frac{\hbar k^{2}}{2m_{LP}}
\left(\frac{\hbar k^2}{2m_{LP}}+2\,\mu \right)}.
\label{bog_eq}
\end{equation}
The non-equilibrium nature of the system is quantified by the effective
relaxation rate 
\begin{equation}
\Gamma=\alpha\, \beta\, \gamma/(1+\alpha\beta),  \label{Gamma}
\end{equation}
whose value tends to $0$ when the threshold is approached
$\alpha\gtrsim 0$ and saturates to $\gamma$ for large $\alpha\gg 1$.
The first $+$ branch of (\ref{spectr}) is the Goldstone branch which
corresponds for small $k$ values to a slow rotation of the condensate
phase, while the $-$ branch corresponds to slow modulations of the
condensate density; for low $k$ values, these are damped at the finite
rate $\Gamma$ defined in (\ref{Gamma}).  From (\ref{spectr}) it is
immediate to obtain a prediction for the width $\Delta k$ of the
region in $k$ space in which the dispersion of the Goldstone mode is
flat $\mathrm{Re}[\omega_G(k)]=0$. Outside this region, i.e. for $k\gg
\Delta k$, the $\pm$ modes recover the standard Bogoliubov modes of an
equilibrium condensate. In order for a sound-like branch to be
observable, one however needs that $\Delta k \ll
1/\xi$, $\xi=\sqrt{\hbar/(m_{LP}\,\mu)}$ being the healing length of
the condensate. Elementary manipulations show that this condition is
actually met if $\gamma^2\ll g\,n_R^0\,(1+\alpha\beta)\,\gamma_R/\beta$.

More complex features are observed in the case where $\gamma_R$ and
$\gamma$ have comparable magnitudes~\cite{note3} and the reservoir can
no longer be adiabatically eliminated [Fig.\ref{fig:disp}(de)].  The
most significant feature is the dynamical instability ${\rm
  Im}[\omega]>0$ of the $DI$ branch that is visible in panel (e) for
$k\xi\lesssim 1.2$: the homogeneous state
(\ref{eq:stat1}-\ref{eq:stat2}) is no longer dynamically stable and a
spatial modulation has to appear in the steady-state density profile
of the condensate.  The origin of the instability can be traced back
in the repulsive interaction between the condensate and the reservoir
polaritons: because of the mean-field interaction potential
${\tilde g} n_R$, a local depletion of the reservoir density $n_R(x)$
creates a potential well which attracts the condensate polaritons
making the local reservoir density drop even more.  This scenario is
only possible for sufficiently slow reservoir relaxation rates $\gamma
_{R}<2\gamma/\left( 1+\alpha \beta \right)$, so that the effect of
repulsive interactions between the reservoir and the condensate is
able to overcome the tendency of density modulations to relax down as
previously discussed.

As a final point of the Letter, we briefly address the effect of the
non-equilibrium condition on the Josephson oscillations between a pair of
spatially separated polariton condensates trapped in adjacent wells.
Following classical work on the Josephson effect~\cite{bec-book,Josephson,smerzi},
 the dynamics can be described by the following set of equations 
\begin{eqnarray}
i \frac{d\, \psi_{j}}{d t}\!\!\!&=&\!\!\!-J\,\psi_{3-j}+ U\,|\psi_{j}|^2\,\psi_{j}
+\frac{i}{2}\big[R(n_j)-\gamma\big]\psi_j  \label{JospehsonPsi} \\
\frac{d\, n_{j}}{dt}\!\!\!&=&\!\!\!P_j-\gamma_R\,n_j-R(n_j)|\psi_j|^2  \label{JosephsonN}
\end{eqnarray}
for the amplitude $\psi_j$ of the two condensates ($j=\{1,2\}$), and the
reservoir densities $n_j$. These equations can be derived by
projecting the GPE onto the localized wavefunctions $\phi_{j}$ describing 
the ground state of each well, and assuming these not to be distorted 
by the interactions. The total
polariton wavefunction then §reads: 
$\psi(\mathbf{r})=\psi_{1}\,\phi_1(\mathbf{r})+\psi_2\,\phi_2(\mathbf{r})$,
the charging energy
$U=g \int\! d\mathbf{r}\,|\phi_j|^4$, and 
the hopping energy is related to the polariton flux through a surface
separating the wells by~\cite{bec-book,note}:
\begin{equation}
J=\frac{\hbar}{m_{LP}}\int_{z=0}\!d\sigma\,(\phi_1^*\,\partial_z\phi_2-\phi_2^*\,\partial_z\phi_1).
\label{hopping}
\end{equation}
The standard normalization $\int\!d\mathbf{r}\,|\phi_j|^2=1$ has been
assumed here.  Inserting typical values of $2\,\mu\mathrm{m}$ for the
well size, a separation between the wells of $1\,\mu{\rm m}$, and a
well depth of $3\,{\rm meV}$, one obtains that $J=0.1\,{\rm meV}$ and
$U=0.03\,{\rm meV}$ for a polariton nonlinearity $g=0.015\, {\rm
  meV}\,\mu{\rm m}^2$.
\begin{figure}[htbp]
\begin{center}
\includegraphics[width=1\columnwidth,angle=0,clip]{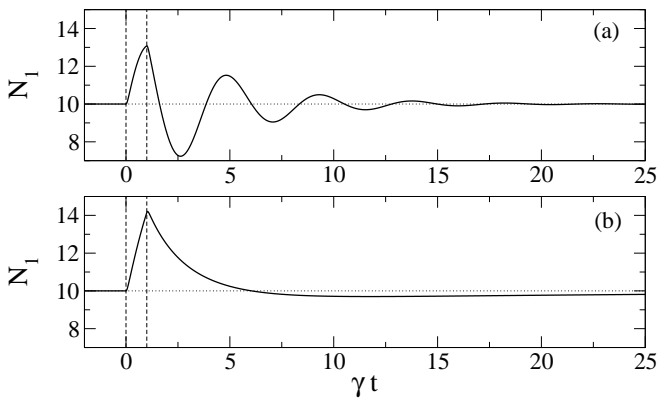}
\end{center}
\caption{Time evolution of the population $N_1$ after an excitation sequence of duration $\gamma T_{exc}=1$ (dashed vertical lines). Upper (a) panel: Josephson oscillations ($J=\gamma/2$). Lower (b) panel: overdamped relaxation ($J=\gamma/20$).
Other parameters: $U=J/10$, $\gamma_R=10\,\gamma$, $R_0=\gamma$, $P_0/P_{thr}=2$, $\Delta P/P_0=0.5$.
}
\label{fig:Josephson}
\end{figure}

Restricting again to the most significant $\gamma_R\gg \gamma$ case,
the frequency of the small amplitude Josephson oscillations around the
stationary state with $N_{1,2}=|\psi_{1,2}|^2=N$ atoms per well can be
obtained by simply replacing the expression
$\omega_{J}=\sqrt{4J(NU+J)}$ to the Bogoliubov frequency
$\omega_{Bog}(k)$ in (\ref{spectr}).  Examples of the different
regimes for the dynamics around the stationary state are shown in
Fig.\ref{fig:Josephson}: starting from the steady-state under
$P_{1,2}=P_0$, the pumping intensity in each well is modulated to
$P_{1,2}=P_0\pm \Delta P$ for a short time interval $0<t<T_{exc}$ and
then brought back to $P_{1,2}=P_0$.  The system dynamics is followed
on the mode populations $N_{1,2}$.  If the effective damping rate
$\Gamma$ is smaller than $\omega_J$ (upper panel), the behaviour
closely ressembles the usual Josephson oscillations as observed in
atomic Bose-Einstein condensates in~\cite{oberthaler}: the only
differences are the intrinsic damping rate $\Gamma$, and the slight
frequency shift predicted by (\ref{spectr}).  On the other hand, if
$\Gamma>\omega_J$ (lower panel), Josephson oscillations are replaced
by an exponential relaxation back to the stationary state; the two
modes at $\omega_\pm$ predicted in (\ref{spectr}) appear in the
relaxation dynamics as two well-separated exponentials.

In summary, we have developped in this Letter a generic model for
the coherent dynamics of a polariton Bose-Einstein condensate.  The
intrinsic non-equilibrium nature of the system is taken into account
by means of a generalized Gross-Pitaevskii equation including loss and
amplification terms.  This model is used to study the elementary
excitations around the stationary state of the system.  In a spatially
homogeneous geometry, a diffusive Goldstone mode is found, and novel
features are predicted for the Josephson effect in two-well
systems. This formalism will be of great utility in the study of the
complex structures which appear in spatially inhomogeneous condensates
thanks to the interplay of condensation and losses (see, e.g.,~\cite{maxime}).

Continuous stimulating discussions with A. Baas, C. Ciuti, M. Richard,
D. Sarchi, and  
V. Savona are warmly acknowledged. This research has been supported
financially  
by the FWO-V projects Nos.G.0356.06, G.0115.06 and the Special
Research Fund  
of the University of Antwerp, BOF NOI UA 2004.M.W. acknowledges
financial support from the  
FWO-Vlaanderen in the form of a ``mandaat  Postdoctoraal Onderzoeker''.


\begin{thebibliography}{99}


\bibitem{Cu2O} J. P. Wolfe, J. L. Lin and D.W. Snoke, in {\em Bose-Einstein
Condensation}, eds. A. Griffin, D. W. Snoke, S. Stringari 
(Cambridge University press, Cambridge, 1995) p. 281. 

\bibitem{CQW} L. V. Butov \emph{et al.}, Nature \textbf{417}, 47 (2002);  D.
W. Snoke \emph{et al.}, Nature \textbf{418}, 754 (2002); Z. V\"oros  
\emph{et al.}, Phys. Rev. Lett. \textbf{94}, 226401 (2005).

\bibitem{kasprzak} J. Kasprzak \emph{et al.}, Nature \textbf{443}, 409
(2006).


\bibitem{Yamamoto} H. Deng \emph{et al.}, Phys. Rev. Lett. \textbf{97}, 
146402 (2006).

\bibitem{GaN} S. Christopoulos {\em et al.} (unpublished).


\bibitem{littlewood} M. H. Szyma\'{n}ska, J. Keeling, and P. B. Littlewood,
Phys. Rev. Lett. \textbf{96}, 230602 (2006).



\bibitem{cross} M.C. Cross and P.C. Hohenberg, Rev. Mod. Phys. \textbf{65},
851 (1993).

\bibitem{goldstone} M. Wouters and I. Carusotto, cond-mat/0606755.


\bibitem{laser} W. Lamb, Phys. Rev. \textbf{134}, A1429 (1964).

\bibitem{atomlaser} B. Kneer, T. Wong, K. Vogel, W. P. Schleich and D.F.
Walls, Phys. Rev. A \textbf{58}, 4841 (1998).


\bibitem{Josephson} G.J. Milburn {\em et al.}, Phys. Rev. A {\bf 55}, 4318 (1997)

\bibitem{smerzi}
A. Smerzi {\em et al.}, Phys. Rev. Lett.{\bf 79}, 4950 (1997); 
S. Giovanazzi {\em et al.}, Phys. Rev. Lett. {\bf 84}, 4521 (2000).

\bibitem{oberthaler} M. Albiez {\em et al.}, Phys. Rev. Lett. {\bf 95}, 010402 (2005).


\bibitem{tassone_bottleneck} F. Tassone, C. Piermarocchi, V. Savona,  A.
Quattropani, and P. Schwendimann, Phys. Rev. B \textbf{56}, 7554  (1997).

\bibitem{porras} D. Porras, C. Ciuti, J. J. Baumberg, and C. Tejedor,  Phys.
Rev. B \textbf{66}, 085304 (2002).

\bibitem{note1} The finite-size of the system guarantees robustness of  the condensate
against long-distance phase  fluctuations~\cite{bec-book}.


\bibitem{4pani} P. Schwendimann and A. Quattropani, Phys. Rev. B \textbf{\ 74}, 045324 (2006).

\bibitem{otherThPolBEC} Y. G. Rubo, F. P. Laussy, G. Malpuech,  A. Kavokin,
and P. Bigenwald, Phys. Rev. Lett. \textbf{91}, 156403  (2003); F. P.
Laussy, G. Malpuech, A. Kavokin, and  P. Bigenwald, Phys. Rev. Lett. 
\textbf{93}, 016402 (2004).

\bibitem{tassone} F. Tassone and Y. Yamamoto, Phys. Rev. B \textbf{59}, 
10830 (1999).

\bibitem{haug} T. D. Doan, H. T. Cao, D. B. Tran Thoai, and H. Haug,  Phys.
Rev. B \textbf{72}, 085301 (2005).

\bibitem{bec-book} L.P. Pitaevskii and S. Stringari, \textsl{Bose-Einstein
Condensation}, Clarendon Press Oxford (2003).

\bibitem{PRL_superfl} I. Carusotto and C. Ciuti, Phys. Rev. Lett. \textbf{\
93}, 166401 (2004).


\bibitem{portella} M.T. Portella-Oberli \emph{et al. }Phys. Rev. B 
\textbf{66}, 155305 (2002).

\bibitem{note2}
Mathematically, it is interesting to observe that the singularity at
$k=0$ in the sound mode dispersion of equilibrium condensates is
intimately related to the fact that this is a ``special
point''~\cite{kato} where the Bogoliubov matrix (equal to the upper-left
$2\times2$ block of $\mathcal{L}_{k}$) is not 
diagonalizable~\cite{phase_diff_bec}.
This property is actually spoiled in the non-equilibrium case by the
coupling to the reservoir.


\bibitem{phase_diff_bec} Y. Castin and R. Dum, Phys. Rev. A \textbf{57},
3008 (1998).

\bibitem{note3} 
Although we do not expect it to be observed in current experiments,
this regime is not in contradiction with the assumption of a 
  local equilibrium within the reservoir: the
  effective decay rate $\gamma_R$ of the active polariton density
  $n_R$ does not necessarily coincide with the equilibration rate of
  the reservoir.


\bibitem{kato} T. Kato, \emph{Perturbation theory for linear operators}
(Sprin\-ger-Ver\-lag, 1984).


\bibitem{note} See~\cite{smerzi} for alternative forms in terms of overlap integrals.



\bibitem{maxime} M. Richard {\em et al.}, Phys. Rev. Lett. {\bf 94}, 
187401 (2005).

\end{thebibliography}
\end{document}